\documentstyle[prd,preprint,tighten,aps,eqsecnum,%
amssymb,newlfont,psfig]{revtex}
\newcommand{\be}{\begin{equation}}
\newcommand{\ee}{\end{equation}}
\newcommand{\ba}{\begin{eqnarray}}
\newcommand{\ea}{\end{eqnarray}}
\newcommand{\no}{\nonumber}
\begin{document}
\draft
\preprint{
\begin{tabular}{r}
UWThPh-1999-14\\
PRL-TH-1999\\
April 1999
\end{tabular}
}
\title{Neutrino oscillations and the effect of the finite lifetime\\
of the  neutrino source \\ 
}
\author{W. Grimus}
\address{Institute for Theoretical Physics, University of Vienna\\         
Boltzmanngasse 5, A-1090, Vienna, Austria}
\author{S. Mohanty and P. Stockinger}
\address{Theory Group, Physical Research Laboratory\\
Ahmedabad - 380 009, India}
\maketitle
\begin{abstract}
We consider a neutrino source at rest and discuss a condition for the 
existence of neutrino oscillations which derives from the finite lifetime
$\tau_S$ of the neutrino source particle. This condition is present if 
the neutrino source is a free particle such that its wave function is 
non-stationary. For a Gaussian wave function and with some 
simplifying assumptions, we  study  the modification 
of the usual oscillation probability stemming from $\tau_S$. In the
present accelerator experiments  the effect of $\tau_S$ can be neglected.
We discuss some experimental situations where the source lifetime becomes
relevant in the oscillation formula. 
\end{abstract}


\narrowtext
\section{Introduction}
\label{introduction}

Neutrino physics is one of the most active fields in particle
physics nowadays. Apart from the impressive results of the
underground experiments concerning atmospheric and solar
neutrino measurements, also reactor and accelerator physics have
a large share in the evolution of this field and the
understanding of the neutrino mass spectrum and mixing matrix
(for recent reviews see, e.g., Refs. \cite{conrad,BGG98}). Since
the phenomenon of neutrino oscillations is now close to being an
established physical reality and constitutes the most tangible
window for physics beyond the Standard Model, it is very important
to know if there are any limitations to the validity of the usual 
formula for neutrino survival and
transition probabilities \cite{pon} with which the experimental results
are evaluated. Such questions have been discussed extensively in
the literature in the context of the wave packet and
field-theoretical approaches (see, e.g., Ref. \cite{zralek} for a list of
references). 

In this paper we use the field-theoretical approach in the spirit of
Ref. \cite{rich},
which has proved to be a general and unambiguous method to analyse
neutrino oscillations, and concentrate on the effect of the finite
lifetime of the neutrino source particle.
We study the problem under the following three assumptions:
\begin{enumerate}
\renewcommand{\labelenumi}{\roman{enumi}.}
\item the neutrino source is at rest,
\item the source particle is not in a bound state and, therefore,
 is described by a non-stationary wave function,
\item the wave function of the detector particle is stationary.
\end{enumerate}
The first assumption is of technical nature and allows us to use
the methods and results of Refs. \cite{GS96,GSM99} and can 
probably be overcome \cite{campagne}, however, the second one 
is essential to the discussion in this paper. 
The third assumption simply means that the particle with which 
the neutrino reacts in the detector is in a bound
state and the wave function of the detector particle does not spread with
time \cite{GSM99}. Note that we also neglect for both the production
and the detection process any possible interaction with the background.

The finite lifetime $\tau_S = 1/\Gamma$ of the neutrino 
source particle has  an impact on the neutrino oscillation
probability in the following  two ways:
\begin{enumerate}
\renewcommand{\labelenumi}{(\alph{enumi})}

\item a suppression of the probability 
amplitude as a function of $\Gamma$ \cite{GSM99}, an effect
independent of $L$, the distance between neutrino source and detection, 

\item a damping of neutrino oscillations through the coherence 
length $L^\mathrm{coh}_\Gamma$ caused by the finite lifetime and 
given by the factor $\exp (-L/L^\mathrm{coh}_\Gamma)$ \cite{GSM99} in 
the oscillation probability. 
 
\end{enumerate}
The main topic of this paper is the derivation of a correction 
to the usual neutrino oscillation probability according to point 
(a) and a numerical study of this correction. It has been
shown in Ref. \cite{GSM99} that the effect (a) of $\Gamma$
is only present provided assumption ii) is valid.
This seems to be the case in the important experiments of the  
LSND \cite{LSND} and KARMEN \cite{KARMEN} Collaborations,
which study  $\bar\nu_\mu\to\bar\nu_e$ oscillations with
$\mu^+$ decay at rest as $\bar\nu_\mu$ source.\footnote{In the LSND
experiment also $\nu_\mu\to\nu_e$ oscillations are investigated where
the muon neutrinos originate from $\pi^+$ decay in flight. According
to assumption i) we do not discuss this case here.}  In the 
following we will use the source and detection reactions of LSND 
and KARMEN as a model for our field-theoretical treatment of 
neutrino oscillations. 

The paper is organized as follows.
In Section II we derive the transition probability 
for $\bar\nu_\mu\rightarrow\bar\nu_e$ oscillations and show how the 
finite source lifetime changes the standard formula of 
neutrino oscillations and gives rise to a suppression factor for the 
interference terms in the oscillation formula. We will see that the 
corresponding condition for the existence of neutrino oscillations,
i.e., the condition that the suppression does not take place, depends 
only on parameters supplied, at least in principle, by the experimental
set-up and the neutrino mass squared difference $\Delta m^2$. This is 
natural within the field-theoretical treatment which enables the 
study of the dependence on those quantities which are really observed 
or manipulated in oscillation experiments.
In Section III we apply the
results of Section II to the case of a Gaussian $\mu^+$ wave
function and make a numerical study of the 2-flavour transition
probability to investigate quantitatively the influence of the
correction factor. In Section IV we present a discussion of our
findings and draw the conclusions.

\section{The cross section}

As mentioned in the introduction, for definiteness and also for
comparison with the notation in Ref. \cite{GSM99} we discuss the process
\be\label{process}
\mu^+ \to e^+ + \nu_e + \bar\nu_\mu 
\stackrel{\nu\: \mathrm{osc.}}{\leadsto}
\bar\nu_e + p \to n + e^+ \,,
\ee
which is investigated in the LSND and KARMEN experiments.
As shown in Ref. \cite{GSM99}, with the help of the Weisskopf--Wigner
approximation one can write the amplitude of 
the process (\ref{process}) in the limit $t \rightarrow \infty$ as
\ba
{\cal A} & = & (-i)^2 
\langle \nu_e (p'_\nu), e^+_S (p'_{eS}); e^+_D (p'_{eD}), n (p'_n) | \no\\
&& \times T \left[ \int_0^\infty \! dt_1 \int d^3x_1 
\int_0^\infty \! dt_2 \int d^3x_2\, {\cal H}^+_{S,\mathrm{int}}(x_1) 
e^{-\frac{1}{2}\Gamma t_1} {\cal H}^+_{D,\mathrm{int}}(x_2) 
\right] |\mu^+ ; p \rangle \,,
\label{a}
\ea
where $T$ is the time-ordering symbol and $\Gamma$ is the total decay width
of the muon. ${\cal H}^+_{S,\mathrm{int}}$ and 
${\cal H}^+_{D,\mathrm{int}}$ 
are the relevant Hamiltonian densities (in the interaction picture) 
for the production and 
the detection of the neutrinos, describing muon decay and proton to neutron
transition (\ref{process}), respectively.
The indices $S$ and $D$ denote source and detection, respectively.
The muon $\mu^+$ and the proton $p$ are localized at the 
coordinates $\vec x_S$ and $\vec x_D$, respectively. 
The proton state is stationary whereas the decaying muon 
at rest is described by a free wave packet with an average momentum equal 
to zero according to the assumptions i)--iii). Hence the 
spinors of the initial particles are written as 
\ba
\Psi_p (x) = \psi_p (\vec x - \vec x_D) \, e^{-i E_p t}
& \quad \mbox{with} \quad &
\psi_p (\vec x-\vec x_D) = \frac{1}{(2\pi)^{3/2}} \int d^3p\,
\widetilde{\psi}_p(\vec p)\, e^{i\vec p \cdot (\vec x-\vec x_D)}
\no\\
& \quad \mbox{and} \quad &
\widetilde{\psi}_p(\vec p) = \widetilde{\psi}^\prime_p(\vec p)
 u_p(\vec p)
\label{proton}
\ea
for the proton and
\ba
\psi_\mu (x) = \int \frac{d^3 p}{(2\pi)^{3/2}}\, \widetilde{\psi}_\mu 
(\vec p)\, e^{- i (\vec p \cdot \vec x - E_\mu(\vec p)t)} \times
e^{i \vec p \cdot \vec x_S} 
& \quad \mbox{with} \quad &
\widetilde{\psi}_\mu(\vec p) = \widetilde{\psi}^\prime_\mu(\vec p)
 v_\mu(\vec p)
\ea
for the muon with $E_\mu(\vec{p}) = \sqrt{m^2_\mu+\vec{p}\,^2}$. 
In the 4-spinors $u_p(\vec p)$ and $v_\mu(\vec p)$
we have left out the polarizations of the proton and muon,
respectively, because they are irrelevant in the further discussion.

The function $\psi_p (\vec y)$ is peaked at $\vec y = \vec 0$
and the wave packet $\widetilde{\psi}_\mu (\vec p)$ 
in momentum space is peaked 
around the average momentum $\langle \vec p\rangle = \vec 0$. 
The final particles are described by plane waves.
After carrying out all integrations, the leading term of the
amplitude in the asymptotic limit $L \to \infty$ 
can be written as \cite{GSM99}
\be\label{ampinfty}
{\cal A}^\infty =   \sum_j U_{\mu j}U^*_{ej} e^{i q_j L} 
{\cal A}^S_j {\cal A}^D_j \,,
\ee
where ${\cal A}^S_j$ and ${\cal A}^D_j$ denote the amplitudes for 
production and detection (\ref{process}) of a neutrino with mass
$m_j$, respectively. The parts of these amplitudes which are important for the
further discussion are given by
\be\label{as}
{\cal A}^S_j =
\frac {1}{i (E_{Sj} -E_D) + \frac{1}{2}\Gamma}\,
\overline{\widetilde{\psi}}_\mu (\vec p_1 + q_j \vec l\,) \,\cdots
\ee
and
\be\label{ad}
{\cal A}^D_j = \cdots\,
\widetilde{\psi}_p(-q_j \vec l  + \vec p_2) \,.
\ee
For the full expressions see Ref. \cite{GSM99}.
The kinematical quantities occurring in Eqs.~(\ref{as}) and (\ref{ad})
are defined by
\ba\label{def1}
q_j & = & \sqrt{E_D^2 - m_j^2} \,, \no\\
E_D & = & E'_n + E'_{eD} - E_p \,, \no\\
E_{Sj}& = & E_\mu (q_j \vec l + \vec p_1) - E'_\nu - E'_{eS}
\ea
and
\be\label{def2}
\vec p_1 = \vec p\,'\!\!_\nu + \vec p\,'\!\!_{eS} \,,\quad
\vec p_2 = \vec p\,'\!\!_n + \vec p\,'\!\!_{eD} 
\quad \mbox{and} \quad  
L = | \vec x_D - \vec x_S | \,.
\ee
Note that the first two formulas in Eq.~(\ref{def1}) follow from
assumption iii) in the introduction \cite{GS96,GSM99}.
Eq.~(\ref{as}) shows the dependence of ${\cal A}^\infty$ on $\Gamma$
(see point (a) in the introduction).
The structure (\ref{ampinfty}) arises from the fact that in 
the limit $L\rightarrow\infty$ the neutrinos with mass $m_j$, described
by an inner line of the Feynman diagram derived from ${\cal A}$
(\ref{a}), are on mass shell. Consequently, the 
amplitude for neutrino production and its subsequent detection factorizes 
for each $j$ into a product of production amplitude and detection amplitude
\cite{GS96}.
Looking at Eqs.~(\ref{ampinfty}), (\ref{as}) and
(\ref{ad}) it is evident that oscillations
involving $m^2_j-m^2_k$ can only take place if the 
conditions \cite{rich,GS96}
\be\label{ACC}
|q_j-q_k| \lesssim \sigma_S \quad \mbox{and} \quad
|q_j-q_k| \lesssim \sigma_D 
\ee
and \cite{GSM99}
\be\label{SFC}
|E_{Sj}-E_{Sk}| \lesssim \frac{1}{2} \Gamma 
\ee
hold, where $\sigma_S$ and $\sigma_D$ are the widths of
$\widetilde{\psi}_\mu$ and $\widetilde{\psi}_p$, respectively.
We call conditions (\ref{ACC}) amplitude coherence conditions (ACC)
and equation (\ref{SFC}) the source wave packet -- finite lifetime 
condition (SFC) \cite{GSM99}. 
If they  are not fulfilled then some terms of the amplitude and 
consequently the corresponding
interference terms in the cross section (oscillation probability) 
are suppressed. 

The cross section $\sigma$ is obtained by taking the absolute 
square of the amplitude (\ref{ampinfty}),
integrating over the final state momenta $\vec p\,'\!\!_\nu$,
$\vec p\,'\!\!_{eS}$,
$\vec p\,'\!\!_n$, $\vec p\,'\!\!_{eD}$ and averaging over the spins
of the initial muon and proton. Hence
\be\label{sigma}
\sigma = \int dP\; |{\cal A}^\infty|^2 \,,
\ee
where $dP$ denotes the integration over the momenta of the particles 
at the external legs and is given by
\be
dP = \frac{d^3p'_\nu}{2E'_\nu}\; \frac{d^3p'_{eS}}{2E'_{eS}}\;
     \frac{d^3p'_n}{2E'_n}\;  \frac{d^3p'_{eD}}{2E'_{eD}}
\ee
and the integration is done over some volume of the phase space.

In general, the integrations in (\ref{sigma})
cannot be performed without knowledge of the source and detector wave 
functions. However, as noticed in Ref. \cite{GSM99}, for
\be\label{IC}
\Gamma \ll \sigma_{S,D} 
\ee
the factors 
\be\label{factor1}
\left\{(i(E_{Sj}-E_D)+\Gamma/2) \times 
(-i(E_{Sk}-E_D)+\Gamma/2)\right\}^{-1}
\ee
in the cross section are strongly peaked with respect to $E_D$
and we interpret the conditions (\ref{IC}) that the rest of the
cross section is flat with respect to $E_D$ if varied over intervals several
orders of magnitude larger than $\Gamma$. This is a reasonable
assumption because in the LSND and KARMEN experiments the
stopped muons have momenta of the order 0.01 MeV \cite{louis,drexlin}, and 
thus $\sigma_S$ will be in the same range. Assuming
atomic dimensions of the spread of the detector particle wave
function, we find $\sigma_D \sim 10^{-3}$ MeV. In any case, even
if our guesses for $\sigma_S$ and $\sigma_D$ are wrong by
several orders of magnitude, these widths can never be so small
like the decay constant of the muon $\Gamma \simeq 3 \times
10^{-16}$ MeV. Therefore, we adopt the procedure that
integrating over momenta of the final state of the detector 
leads to an integration in the variable $E_D$ and in view of the
expressions (\ref{factor1}) it suffices to pick out an
integration interval of length $\Delta E_D$ which fulfills
$\Gamma \ll \Delta E_D \ll \sigma_{S,D}$. Considering neutrino
oscillations with respect to $m^2_j - m^2_k \equiv \Delta m^2_{jk}$, 
then this integration interval should contain 
$E_{Sj}$ and $E_{Sk}$. Since the condition (\ref{SFC}) is
necessary for these oscillations to happen at all, we define a
mean value 
\be\label{dED}
\bar{E}_D \equiv \langle E_{Sj} \rangle
\ee
for the neutrino mass eigenfields participating in the
oscillations and $\Delta E_D$ only has to be a few orders of
magnitude larger than $\Gamma$. In addition, the expressions
(\ref{factor1}) suggest to close the integration over the
interval of length $\Delta E_D$ along the real axis by a half
circle in the complex plane and to apply Cauchy's theorem \cite{GSM99}.

In this way we have shown that the cross section contains the
damping factor \cite{GSM99}
\be\label{damping}
\exp \left( - \frac{\Delta m^2_{jk}\Gamma}{4\bar{E}^2_D}\,L \right) 
\ee
leading to a suppression 
of the corresponding interference term in the oscillation
probability when $L$ becomes greater than the coherence length
$L^\mathrm{coh}_\Gamma = 4 \bar E_D^2/\Delta m^2_{jk} \Gamma$. 
As already discussed in Ref. 
\cite{GSM99}, this coherence length is of the order of 100 light years 
for typical input relevant for the LSND and KARMEN experiments
and its effect is certainly negligible for all experiments 
with terrestrial neutrinos. In the following computations we 
will neglect the coherence length arising from the finite lifetime of the
neutrino source.

Then, the only relevant integrand with respect to $E_D$ is
given by the factors (\ref{factor1}) and we get
\be\label{int1}
\int_{-\infty}^\infty dE_D\,\frac{1}{i(E_{Sj}-E_D)+\frac{1}{2}\Gamma}\,
\frac{1}{-i(E_{Sk}-E_D)+\frac{1}{2}\Gamma} = \frac{2\pi}{\Gamma}\,\frac{1}
{1+i\frac{1}{\Gamma}(E_{Sj}-E_{Sk})} \,.
\ee
On the right-hand side of this equation we can use assumption
i) (the muon at rest) to compute to a very good approxmimation 
the energy difference
\be
E_{Sj}-E_{Sk}\simeq -\frac{\Delta m^2_{jk}}{2m_\mu \bar E_D}\,\vec{l}
\!\cdot\!(\vec{p_1}
+\bar E_D \vec{l}\,) \,.
\ee
As the next step, we want to perform the integration over 
$d^3p'_\nu$. Note that the momentum $\vec p\,'\!\!_\nu$ of
the $\nu_e$ from the source reaction (see Eq.~(\ref{process})) 
cannot be measured. 
Again we make a simplifying approximation based on assumption i), 
namely that the muon wave function can be represented by
\be
\widetilde{\psi}_\mu(\vec{p}_1+q_j\vec{l}\,) \simeq 
\widetilde{\psi}'_\mu(\vec{p}_1+\bar E_D \vec{l}\,)\,
v_\mu(\vec{0}\,) \,.
\ee
Then the integration over 
$d^3p'_\nu$ concerns only the functions $(E_{Sj}-E_{Sk})$ 
in (\ref{int1}) and 
$\widetilde{\psi}^\prime_\mu$. Therefore, we define the quantities
\be\label{gfact}
g_{jk}= \int d^3u\, |\widetilde{\psi}'_\mu(\vec{u})|^2 
\frac{1}{1-i\rho_{jk}\,\vec{l}
\!\cdot\! \vec{u}/\sigma_S}\, ,
\ee
where we have defined $\vec{u}\equiv \vec{p}_1+E_\nu\vec{l}$ and
\be\label{rho}
\rho_{jk}\equiv\frac{\Delta m^2_{jk}\sigma_S}{2 m_\mu E_\nu\Gamma} \,.
\ee 
Note that in view of Eqs.~(\ref{def1}) and (\ref{dED}) we identify
the neutrino energy which can in principle be measured by the
detector as
\be\label{eneu}
E_\nu\equiv\bar E_D \,.
\ee
The quantities (\ref{gfact}) have the properties that
$g_{jk}=g_{kj}^*$ and $g_{jj}=1$. Since we consider
ultrarelativistic neutrinos, we can take the limit
$m_j \rightarrow 0$ $\forall \,j$ in all terms of the cross 
section (\ref{sigma}) except the oscillation phases and 
$g_{jk}$. Then the probability for $\bar\nu_\mu\to\bar\nu_e$ 
oscillations is given by
\ba
P_{\bar\nu_\mu\to\bar\nu_e} &=& 
\sum_j |U_{e j}|^2 |U_{\mu j}|^2 +\no\\
&&2\, \mbox{Re} \left\{ \sum_{j>k} 
U^*_{e j} U_{\mu j} U_{e k} U^*_{\mu k}\,\,
g_{jk}\,\exp \left( -i \frac{\Delta m^2_{jk} L}{2E_\nu} \right)\, 
\right\} \,.
\label{P}
\ea
In contrast to the formula derived within the standard 
treatment of neutrino oscillations, the quantities $g_{jk}$ appear 
as correction factors in 
Eq.~(\ref{P}). These non-standard factors $g_{jk}$
represent the quantitative effect of the SFC condition (\ref{SFC})
discussed in Ref. \cite{GSM99}. They arise from assumption ii)
(see introduction) that the muon is described by a free wave function,
which does not have a sharp energy, and would disappear for a muon
in a stationary (bound) state (see Ref. \cite{GS96}).

\section{The case of a Gaussian muon wave packet}

In order to get a feeling for the effect of the $g$-factors (\ref{gfact})
on the oscillation probabilities we confine 
ourselves now to the case of a Gaussian muon wavefunction
\be\label{gauss}
\widetilde{\psi}^\prime_\mu(\vec p) = (\sqrt{ \pi} \sigma_S)^{-3/2}
 \exp{\left(-\frac{\vec p\,^2}{2\sigma_S^2}\right)} \,.
\ee
Then we get
\be\label{ggfact}
g_{jk}=\frac{1}{\sqrt{\pi}\sigma_S}\int_{-\infty}^\infty du\, 
e^{-u^2/\sigma^2_S} \frac{1}{1-i\rho_{jk} u/\sigma_S} \,,
\ee
where $u\equiv \vec{l}\!\cdot\!\vec{u}$.
Since the exponential factor in Eq.~(\ref{ggfact}) is an even 
function with respect to the variable $u$ the $g$-factors become 
real and defining $y = u/\sigma_S$ one obtains $g_{jk} = g(\rho_{jk})$
with
\be\label{g}
g(\rho) \equiv 
\frac{1}{\sqrt{\pi}}\int_{-\infty}^\infty dy \frac{e^{-y^2}}{1+
\rho^2 y^2} 
=\frac{\sqrt{\pi}}{\rho} \exp \left( \frac{1}{\rho^2} \right)
\Phi_c \left( \frac{1}{\rho} \right) \,,
\ee
where $\Phi_c$ is the complementary error function (see, e.g., Ref. 
\cite{Gradst}). Note that 
$g(\rho) \simeq 1-\rho^2/2$ for $\rho \ll 1$
and $g(\rho) \to \sqrt{\pi}/\rho$ for $\rho \to \infty$. 
Under the assumption of oscillations between two neutrino flavours
the transition probability becomes 
\be\label{gProb}
P_{\bar\nu_\mu\rightarrow\bar\nu_e}=\frac{1}{2}\sin^2 2\theta
\left( 1-g_{12} \cos\frac{\Delta m^2 L}{2 E_\nu} \right) \,,
\ee
where $\theta$ is the 2-flavour mixing angle. In comparison with 
the standard
formula the factor $g_{12}$ in (\ref{gProb}) lowers for a given transition 
probability the upper bounds on $\Delta m^2$ and $\sin^22\theta$.
This can be seen explicitly by considering 
the low $\Delta m^2$-region of the parameter
space where the cosine in  expression (\ref{gProb}) can be 
expanded in $\Delta
m^2$. If now $\rho\lesssim 1$ holds for an average $\Delta m^2$, in the 
low $\Delta m^2$-region we get $\rho\ll 1$ and can use the corresponding
expansion of $g(\rho)$ given above. 
Then, in this region the isoprobability contour is determined 
by the relation
\be
(\Delta m^2 )^{2}\,\sin^2 2\theta \simeq 
{16E_\nu^2\, P_{\bar\nu_\mu\rightarrow\bar\nu_e} \over
L^2 + (\frac{\sigma_S\tau_S}{m_\mu})^2}\, .
\label{isop}
\ee
It is clear from Eq.~(\ref{isop}) that a large source lifetime and a large
source momentum spread  lower the upper bounds on $\Delta m^2$
and $\sin^2 2\theta $ compared to the standard oscillation formula. For the
source lifetime effect to be significant in a given experiment, 
the numerical value of $\sigma_S\tau_S/m_\mu$ must be 
comparable to the source detector distance $L$.

In the following we investigate quantitatively the effect of $g_{12}$
on the transition probability (\ref{gProb}). In Fig.~1 we display  
the function $g(\rho)$. One can see that at $\rho = 0.5$ 
the function $g$ is around 0.9 which means that $g_{12}$
starts to deviate appreciably from 1 when $\rho$ becomes greater than 
about 0.5. In the case of the LSND and KARMEN experiments, using
$\Gamma \simeq 3\times 10^{-16}$ MeV for the decay width of the muon
and the typical numbers $\sigma_S \simeq 0.01$ MeV \cite{louis,drexlin},
$\Delta m^2 \simeq 1$ eV$^2$ and $E_\nu \simeq 30$ MeV, we obtain
$\rho_{12}\simeq 0.5\times 10^{-2}$.
From this estimate we conclude that with the above input numbers
the correction to the transition probability is negligible for
the LSND and KARMEN experiments because
$1-g_{12} \sim 10^{-5}$ (see remark after definition of $g$ (\ref{g})). 
Note, however, that $\rho_{12}\simeq 0.5\times 10^{-2}$ is only
two orders of magnitude away from having a 10\% effect.
This motivates us to have a look at the influence of $\rho_{12}$ ($g_{12}$)
on exclusion curves obtained, for instance, in the KARMEN
experiment.
Therefore, we use $L = 17.7$ m, average in Eq.~(\ref{gProb}) over the energy
spectrum of $\bar{\nu}_\mu$ and as a Gedankenexperiment we vary the momentum
spread $\sigma_S$. 
In Fig.~2 we show the corresponding exclusion curves in the $\Delta m^2$
-- $\sin^2 2\theta$ plane for the cases of
$\sigma_S=0$, 1 and 10 MeV corresponding to $g_{12} \simeq
1$, 0.9 and 0.3, respectively. 
We can see that for $\sigma_S=10$ MeV the exclusion curve has changed
noticeably.

\section{Discussion and conclusions}

In this paper we have studied the effects of the finite lifetime of
the neutrino source particle. In particular, we have eleborated 
the condition for the existence of 
neutrino oscillations arising in a situation where the neutrino
source particle is not described by a bound state and thus
represented by a non-stationary wave function. 
Furthermore, in our analysis we have
made the assumption that the neutrino source is at rest.
In order to perform an analysis closely related to experiment we 
used the field-theoretical approach which allows to work directly with
the states of the particles responsible for neutrino production and 
detection
and where the oscillating neutrinos are represented by an inner line 
in the complete Feynman diagram containing the source and detection 
processes.
Thus the neutrino oscillation amplitude is determined simply through 
the neutrino production and detection interaction Hamiltonians.
For simplicity, in the following we will concentrate on 2-flavour
oscillations. 

The finite lifetime of the neutrino source leads to a finite coherence
length \cite{KNW96}. In our case we have explicitly indicated the
damping factor (\ref{damping}) from where the coherence length 
$L^\mathrm{coh}_\Gamma$ associated with $\Gamma$ can be read
off. Using numbers taken from the LSND and KARMEN experiments for the
quantities appearing in $L^\mathrm{coh}_\Gamma$ we get a length of the
order of 100 light years \cite{GSM99} which is absolutely negligible
compared the the coherence length stemming from the fact that the
neutrino energy $E_D$ (see Eqs.~(\ref{def1}) and (\ref{dED})) 
determined by the energies
of the final states in the detection process is not accurately
known. Therefore, the effect (b) of $\Gamma$ (see introduction) is
irrelevant. 

Let us now compare the amplitude coherence condition (ACC) with the source
wave packet -- finite lifetime condition (SFC). The ACC can be
reformulated as \cite{kayser,rich,GS96}
\be\label{ACC1}
\sigma_{xS,\, xD} \lesssim \frac{1}{4\pi} L^\mathrm{osc} \, ,
\ee
where $\sigma_{xS,\, xD}$ are the widths of the source and detector
wave functions, respectively, in coordinate space and 
$L^\mathrm{osc}=
4\pi E_\nu/\Delta m^2=2.48\; \mbox{m}\; (E_\nu/1\;\mbox{MeV})
(1\;\mbox{eV}^2/\Delta m^2)$ is the oscillation length. 
For the LSND and KARMEN 
experiments with $E_\nu\sim30$ MeV and $\Delta m^2\sim 1\;\mbox{eV}^2$,
as found by the LSND collaboration, 
one gets $L^\mathrm{osc}\sim 75$ m. This is to be compared
with the widths $\sigma_{xS,\, xD}$ which are certainly in the 
range of atomic distances or smaller. Therefore the conditions
(\ref{ACC1}) are very well fulfilled.
We want to stress once more that the SFC and the correction factor $g_{12}$
(\ref{gfact}) corresponding to effect (a) of $\Gamma$ (see
introduction) derive from two ingredients:
from the finite lifetime of the muon taken into account through the 
Weisskopf--Wigner approximation and from
assumption ii) in the introduction that the wave function of the muon
is non-stationary. The first ingredient leads to 
the factor $1/(i(E_{Sj}-E_D)+\frac{1}{2}\Gamma)$ in the amplitude instead 
of the familiar delta function. The second one leads to the momentum
dependence of $E_{Sj}$ as given in Eq.~(\ref{def1}). If the neutrino 
source particle were in a bound state and, therefore, described by a
stationary wave function, $E_S$ would be fixed and independent of $j$
and momenta, and there would be no condition (\ref{SFC}). Note that in 
our approximation the Hamiltonian which determines
the wave function of the source particle does not
include weak interactions. However, the decay of the source particle 
via the weak interactions -- whether its wave function is stationary or
non-stationary -- is explicitly given by the factor 
$\exp (-\Gamma t_1/2)$ in the amplitude (\ref{a}) according to the
Weisskopf--Wigner approximation.
As shown in Section III the correction factor $g_{12}$ depends on 
the quantity
\be\label{rho1}
\rho = \frac{\Delta m^2 \sigma_S}{2 m_\mu E_\nu\Gamma}\; ,
\ee
and is only sizeably different from one if $\rho \gtrsim 1$. Thus
a convenient formulation of SFC is $\rho < 1$. 
Note that by defining a spread in velocity of the source
wave packet by $\Delta v_S = \sigma_S/m_\mu$ the last formulation of
SFC can be rewritten in analogy to Eq.~(\ref{ACC1}) as \cite{GSM99}
\be\label{SFC1}
\Delta v_S \tau_S \lesssim \frac{1}{4\pi} L^\mathrm{osc} \,.
\ee
As we have seen in 
Section III with $\sigma_S \sim 0.01$ MeV one gets $\rho \sim 0.005$
and SFC is well fulfilled for the LSND and KARMEN experiments,
however, the margin for violation of SFC is only two orders of
magnitude, in contrast to ACC where the margin is at least ten orders
of magnitude. 

Taking this observation as a starting point,
we want to see if there is
some chance to increase $\rho$ to 1 or more in order to have an
observable suppression effect.
Looking at Fig.~2 one recognizes that in the case of 
$\sigma_S=10$ MeV the lower line of the curve has undergone a
considerable shift downwards compared to the case $\rho = 0$. 
Note that this shift happens in 
the sensitive region of the LSND and KARMEN experiments. One might
therefore ask the question if it is possible to increase $\sigma_S$ in
order to achieve $\rho \gtrsim 1$. If
the source wave function is such that $\sigma_S\, \sigma_{xS} \sim 1$,
which is correct for Gaussian wave functions,
one would have to localize the neutrino
source particle very well in coordinate space, namely to $\sigma_{xS}
\lesssim 10^{-12}$ m
for $\sigma_S \gtrsim 0.1$ MeV.
Taking a look at Eq.~(\ref{rho1}), one
observes that $\rho$ can also be increased by decreasing the neutrino
energy. However, in this case one lowers the ratio of true over
background events.
Finally,  a small product of mass times decay width of the
source 
particle also enhances $\rho$ but, unfortunately, the muon
has already
the smallest such product among the particles which can
copiously be
produced.\footnote{The corresponding
product for the neutron is smaller, however, to observe this effect in
neutrinos from neutron decay one would have to isolate the neutrons from the
external interactions -- something that seems to be hard to achieve in
experiments.}

The quantum-mechanical ACC and SFC in the forms (\ref{ACC1}) and 
(\ref{SFC1}), respectively, have analogous ``classical
macroscopic'' conditions stemming from the inescapable averaging over some 
regions of the target and the detector, respectively, when an
experiment is performed.
Replacing in Eq.~(\ref{ACC1}) the widths $\sigma_{xS,\, xD}$ 
by the typical sizes $R_{S,D}$, respectively, of the regions in the target 
where the muon is stopped and in the detector where the neutrino detection 
process is localized we arrive at conditions which are obtained by the
incoherent averaging of the oscillation probabilities over the variations
of $L$ due to $R_{S,D}$  and the requirement that 
neutrino oscillations should not be washed out by this averaging process. 
Since $R_{S,D}$ are macroscopic quantities,
if these classical conditions are fulfilled, then clearly also ACC
holds because $\sigma_{xS,\, xD} \ll R_{S,D}$. The classical analogue
to SFC (\ref{SFC1}) says that during its lifetime the 
neutrino source particle should move a distance much 
less than the oscillation length in order not to wash out neutrino
oscillations. This classical condition is not obviously linked to SFC.

\newpage
\begin{figure}
{\setlength{\unitlength} {1mm}
\begin{picture}(155,150)(0,0)
\put(-29,-100){\mbox{\psfig{figure=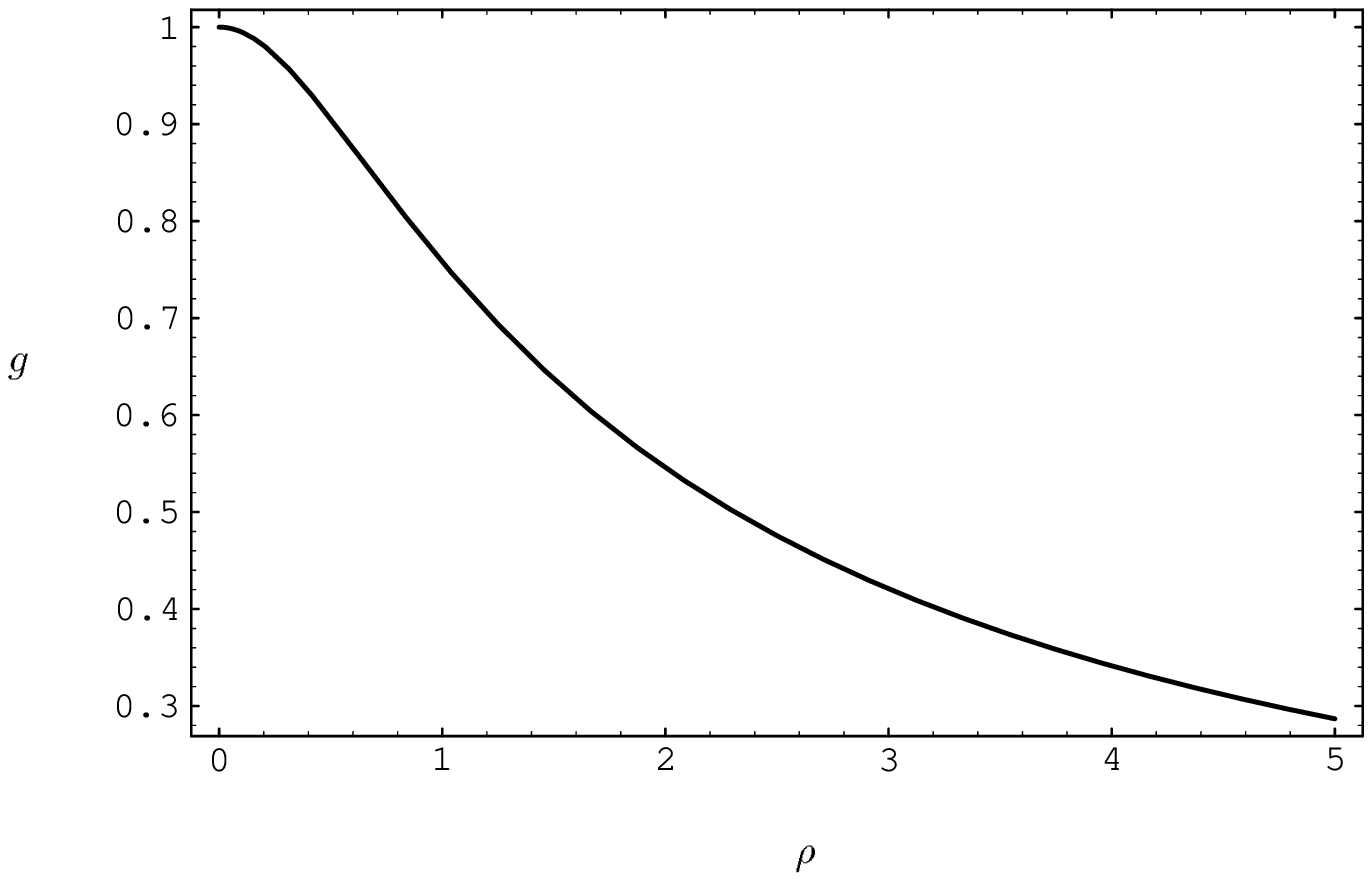,height=30.cm}}}
\end{picture}}
\caption{Plot of the correction factor $g$ (Eq. (\ref{g})) as a
function of $\rho$.}

\end{figure}

\newpage

\begin{figure}
{\setlength{\unitlength} {1mm}
\begin{picture}(155,190)
\put(-17,-40){\mbox{\psfig{figure=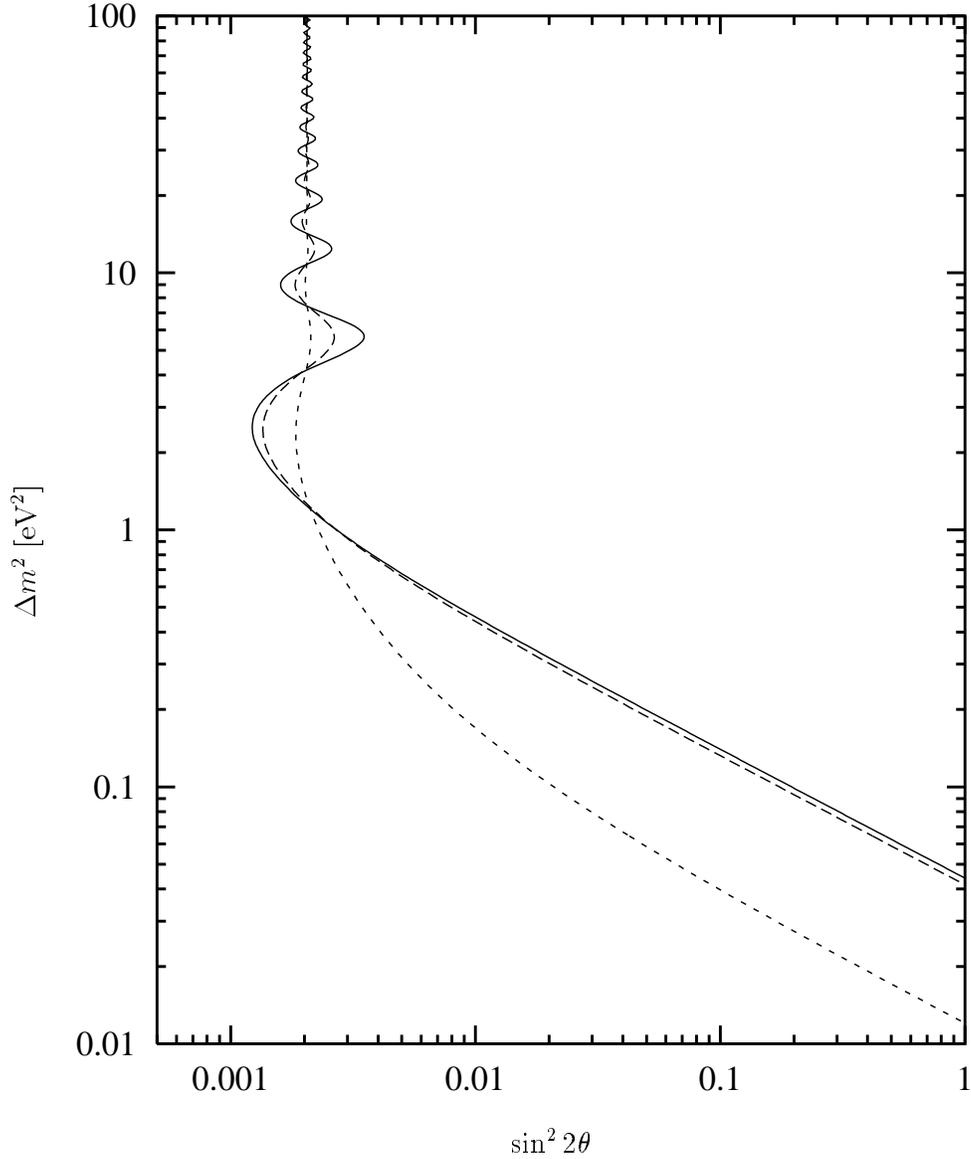,height=28.cm}}}
\end{picture}}
\caption{Contour plot for the transition probability
$\bar P=\int \phi(E_\nu) P_{\bar\nu_\mu\rightarrow\bar\nu_e}
(E_\nu) dE_\nu$ at $\bar P= 0.001$ in the
$\Delta m^2$ -- $\sin^22\theta$ plane where $P_{\bar\nu_\mu\rightarrow
\bar\nu_e}$ is defined in Eq.~(\ref{gProb}) and $\phi$ denotes the
energy distribution of the $\bar\nu_\mu$ flux (see, e.g., Ref. [16]).
We integrated over the neutrino energy $E_\nu$ from   
12 up to 53 MeV and used $L=17.7$ m for the source -- detector
distance $L$. The solid, dashed and  
dotted line correspond to $\sigma_S=0$ ($g=1$), 1 and 10 MeV,
respectively.
}
\end{figure}
\end{document}